# Hierarchy of Periodic patterns in the twist-bend nematic phase of mesogenic dimers

Vitaly P. Panov[1], Michael C. M. Varney[2], Ivan I. Smalyukh[2], Jagdish K. Vij[3], Maria G. Tamba[4] and Georg H. Mehl[4]

1 School of Physics and Astronomy, University of Manchester, Manchester M13 9PL, United Kingdom
2 Department of Physics, University of Colorado, Boulder, USA
3 Department of Electronic and Electrical Engineering, Trinity College Dublin, Dublin 2, Ireland
4 Department of Chemistry, University of Hull, Hull, HU6 7RX, UK

**Abstract.**

Several attractive properties of hydrocarbon-linked mesogenic dimers have been discovered, a major feature of these has been the ability of dimers under confinement to form periodic patterns. At least six different types of stripe-like periodic patterns can be observed in these materials under various experimental conditions. Our most recent finding is the discovery of micrometre-scale periodic patterns at the N-$N_{tb}$ phase transition interface, and sub-micrometre scale patterns in the $N_{tb}$ phase. Layer-like sub-micrometre patterns were observed using three-photon excitation fluorescence polarizing microscopy, and these are being reported here.

**Introduction.**

During recent years, twist-bend nematic phase ($N_{tb}$, or $N_x$ in some early publications) reported in carbon chain linked mesogenic dimers has become a major focal point in liquid crystal research [1, 2, 3]. The potential for applications is enhanced by the existence of the twist-bend nematic phase at room temperature, tuning of its properties by the mixture composition, and its presence in a wide variety of materials such as dimers and a subset of the bent-core molecules [4]. Current scientific interest is generated by a number of highly unusual properties reported for the $N_{tb}$ phase, including the formation of chiral domains by non-chiral molecules [5, 6], ultrafast switching induced by an interaction between flexoelectricity and the short-pitch helical structures [5, 7], and the unusual values of the elastic constants displayed by this phase [8].

A particularly intriguing property of the $N_{tb}$ phase is related to the formation of periodic patterns within its structure. Currently we can highlight at least six different periodicities found in these materials under various experimental conditions. These are listed here as follows.

(I)  The most fundamental periodic structure is a short-scale (~8 nm pitch) tilted helix experimentally observed by Freeze-Fracture Transmission Electron microscopy (FF-TEM) [9, 10]. This finding explains microsecond switching properties of the phase [5, 7] and eliminates a discrepancy between the earlier theoretical predictions and the experimental observations of unusual values of elastic constants [1, 11, 12]. FF-TEM also confirms the absence of smectic layers in the $N_{tb}$ phase as previously found by x-ray experiments [1]. In contrast to the well-established cholesteric liquid crystals, the local director in the $N_{tb}$ phase is tilted by an angle between 9°-37° from the axis of the helix [13].

(II) Another periodic structure is formed by the domains of opposing chirality under the influence of high amplitude AC electric fields [14]. These observations are still awaiting an appropriate theoretical model to be given to facilitating further elucidation and an experimental exploitation of this phenomenon.

(III) Spectacular and colorful periodic patterns observed by Polarizing Optical Microscopy (POM) have attracted wide attention to this class of materials [1], an example of which is the so called "self-deformation stripes" that have a well-defined periodicity controlled by confining surface geometry. In planar cells the periodicity of these stripes is found to be exactly double the confining cell gap (these are set between 2 and 30 micrometers). The most recent studies [15] propose a Helfrich-Hurault type mechanism associated pseudo-layers to explain this self-deformation, although it is not yet clear how this model explains experimentally observed linear dependence of pattern periodicity as a function of the confining cell gap.

(IV) The micrometer-scale periodic patterns (Figure 1) observed at the N-$N_{tb}$ interface [16] are likely to be due to a delicate interplay of the various elastic constants at certain concentrations of dimer-monomer mixtures. The current research in part is focused towards finding a detailed model of this interaction.

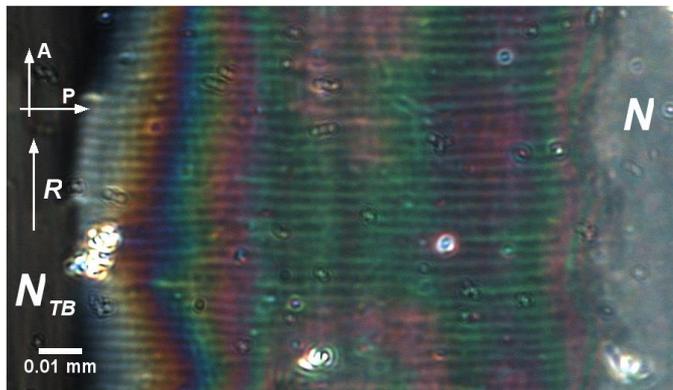

Figure 1. Periodic texture observed upon slow cooling from a nematic phase. Observations made in a 25 μm cell filled with a mixture of CB-C7-CB (70% w/w) and 5CB (30% w/w). The striped region between the N and $N_{tb}$ phase is uniform and have a periodicity of ~ 2.94 μm. [16]

(V) Sub-micrometer fine super-texture has been observed in coexistence with self-deformation patterns described in (III). Although these textures are below the limit of the resolution of an ordinary polarizing microscope, they determine the observed "rope-like" appearance of the self-deformation stripes [2]. Studies performed using Fluorescent Confocal Polarizing Microscopy (FCPM) [17, Figure 2 (d, e)] reveal existence of sub-micron periodic patterns that result from a delicate interplay of material constants which may sporadically exist in the sample depending on factors such as the thermal history of the sample.

(VI) A final characteristic feature of the $N_{tb}$ phase is the formation of focal conic domains through a unique, non-periodic mode and indicated by the presence of a layered structure within the material. This mode often coexists with the self-deformation stripes. This structure is thought to be a reason for an earlier misidentification of this phase to be smectic although such domains in the $N_{tb}$ phase are of different nature than for smectics. In this paper we report an observation of these sub-micron layers in focal conic domains using three-photon excitation fluorescence polarizing microscopy (3PEF-PM).

**Experiment.**

The structure of the material under investigation is shown in Figure 2. A cyano-biphenyl-base dimer with the chain length n=9 (CB-C9-CB) was mixed with a room-temperature nematic 4-Cyano-4'-pentylbiphenyl (5CB) with a weight ratio of 55% CB-C7-CB and 45% 5CB so as to lower the nematic-isotropic transition temperature to that required for the experiment, and to allow the use of the high-resolution optics.

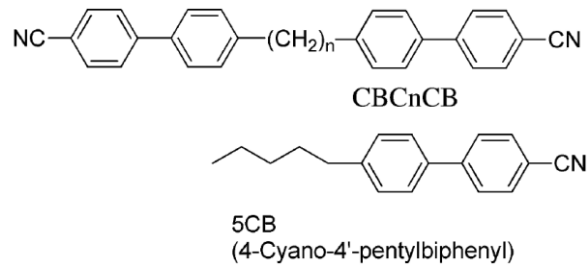

Figure 2. Structure of the material under investigation.

The material studied was infused into a planar rubbed cell constructed from clean glass slide substrates coated with polyimide to set planar anchoring boundary conditions for **n**(r). Cell thickness is approximately 10 μm.

The 3D imaging was performed using three-photon excitation fluorescence polarizing microscopy (3PEF-PM) [18, 19]. This nonlinear optical imaging approach is based on the fluorescence properties of biphenyl groups within certain LC molecules which are excited through three photon absorption of femtosecond infrared laser light. The 3PEF-PM fluorescence intensity exhibits a strong, well-defined dependence [19] on the orientation of linear polarization of the excitation beam relative to the molecular director **n** at the focus of the microscope objective. The resulting images comprise a stack of optical slices which reveal director structures within the studied LC.

Tunable (680-1080 nm) Ti-Sapphire femtosecond oscillator (140 fs, 80 MHz, Chameleon Ultra II, Coherent) is used in home-built microscopy setup (University of Colorado, Boulder, USA), as a laser excitation source for the nonlinear optical imaging in the 3PEF-PM mode. A Faraday isolator is used to protect the Ti:Sapphire laser from back-reflection, and a combination of a half-wave plate and a Glan-laser polarizer allows us to control power and polarization of the laser. The laser pulse is shunted into a scanning mirror unit, which is used in conjunction with a stepper motor for repositioning the focused excitation beam within the sample for three-dimensional point-by-point 3PEF-PM imaging. Software (Fluoview, Olympus, Japan) then allows for construction of three-dimensional images as well as different cross-sections of such images. The excitation light is focused into the sample by use of a high numerical aperture (1.4) objective lens (100x, oil-immersion) and collected by another objective of the same numerical aperture (60x, oil-immersion) in forward mode or the same one for detection in epi-detection mode. The nonlinear signals are detected with photomultiplier tubes (H5784-20, Hamamatsu) within a spectral range of 400-450 nm after passing through a series of dichroic mirrors and bandpass filters. LC self-fluorescence spectra are measured by a spectrometer USB2000-FLG (from Ocean Optics). The technique can achieve 0.2 μm radial and 0.4 μm axial spatial resolution [19].

Figure 3 presents several frames of a 3-D scan of the cell using 3PEF-PM. The non-linear imaging reveals a periodic structure distinct from those previously reported. Fine lines separated from each other by 0.5-2 µm are visible and are running orthogonal to the rubbing direction. Being comparable to the wavelength of the visible light, this periodic structure has been unresolved in the previous observations, but is evident using the high resolution non-linear imaging system. The 3-dimensional scan reveals periodic layers crossing the focal conic pattern.

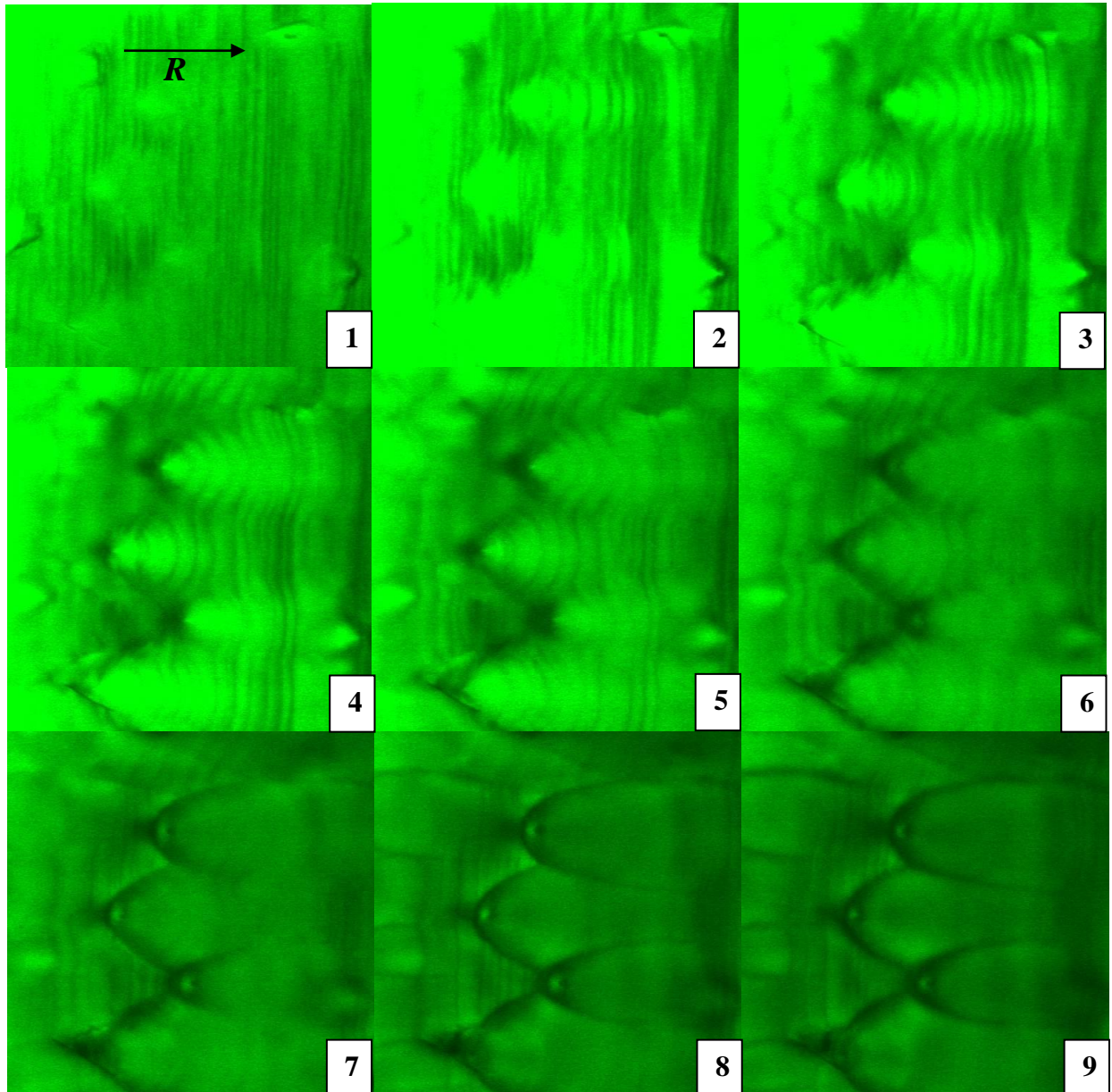

Figure 3. Sliced 3D imaging pattern of a planar cell. Rubbing direction *R* is horizontal and parallel to the incident polarization. Frame size is 36µm, frames separation along the optical axis is 1.2 µm.

**Conclusions.**

In mesogenic dimers and other materials possessing $N_{tb}$ phase one should consider a hierarchy of periodic patterns with characteristic scales ranging from a few molecular lengths (~8 nm) to sample size (units to tens of micrometres). These can lead to a variety of applications including nano-and micro patterning, colloid and particle positioning and manipulation and guided self-assembly of complicated structure. For instance, the sub-micron "layered" pattern reported here may correspond to the pseudo-layers proposed by the Kent group for explaining the self-deformation stripes by Helfrich-Hurault-type mechanism [15]. Since these self-deformation patterns depend on the macroscopic parameters of the sample (such as the confining surface size and alignment), varying these parameters should provide for a rich experimental space.